          \documentclass[preprint, superscriptaddress]{revtex4-1}
\pdfoutput=1
\usepackage{amsmath}%
\usepackage{amsfonts}%
\usepackage{amssymb}%
\usepackage{graphicx}
\usepackage{subfigure}
\newcommand{\beq}{\begin{equation}}
\newcommand{\eeq}{\end{equation}}
\newcommand{\mrm}[1]{\mathrm{#1}}

\begin{document}
\title{Fundamental switching field distribution of a single domain particle derived from the N\'eel-Brown model}

\author{L. Breth}
\email{leoni.breth.fl@ait.ac.at}
\affiliation{AIT - Austrian Institute of Technology, Health and Environment Dept., 1220 Vienna, Austria}
\affiliation{Vienna University of Technology, Solid State Physics Dept., 1040 Vienna, Austria}
\author{D. Suess}
\affiliation{Vienna University of Technology, Solid State Physics Dept., 1040 Vienna, Austria}
\author{C. Vogler}
\affiliation{Vienna University of Technology, Solid State Physics Dept., 1040 Vienna, Austria}
\author{B. Bergmair}
\affiliation{Vienna University of Technology, Solid State Physics Dept., 1040 Vienna, Austria}
\author{M. Fuger}
\affiliation{Vienna University of Technology, Solid State Physics Dept., 1040 Vienna, Austria}
\author{R. Heer}
\affiliation{AIT - Austrian Institute of Technology, Health and Environment Dept., 1220 Vienna, Austria}
\author{H. Brueckl}
\affiliation{AIT - Austrian Institute of Technology, Health and Environment Dept., 1220 Vienna, Austria}

\date{\today}

\begin{abstract}

We present an analytical derivation of the switching field distribution for a single domain particle from the N\'eel-Brown model in the presence of a linearly swept magnetic field and influenced by thermal fluctuations. We show that the switching field distribution corresponds to a probability density function and can be obtained by solving a master equation for the not-switching probability together with the transition rate for the magnetization according to the Arrhenius-N\'eel Law. By calculating the first and second moments of the probability density function we succeed in modeling rate-dependent coercivity and the standard deviation of the coercive field. Complementary to the analytical approach, we also present a Monte Carlo simulation for the switching of a macrospin, which allows us to account for the field dependence of the attempt frequency. The results show excellent agreement with results from a Langevin dynamics simulation and therefore point out the importance to include the relevant dependencies in the attempt frequency. However, we conclude that the N\'eel-Brown model fails to predict switching fields correctly for common field rates and material parameters used in magnetic recording from the loss of normalization of the probability density function. Investigating the transition regime between thermally assisted and dynamic switching will be of future interest regarding the development of new magnetic recording technologies. 

%We present an analytical derivation of the probability density function (pdf) for the switching field distribution of a single domain particle from the master equation for the not-switching probability and the Arrhenius-N\'eel Law. This way, we are able to calculate the mean switching field and its standard deviation in terms of the first and second moments of the pdf. Results of a Monte-Carlo simulation for a swept field experiment show good agreement with the analytical pdf up to a certain field rate, where the pdf becomes unnormalized. We use typical parameters for perpendicular recording media and compare our results to existing models describing the rate dependence of coercivity.
\end{abstract}
\maketitle

\section{Introduction}
The switching behavior of a single domain magnetic particle is strongly influenced by the presence of thermal fluctuations, which help the magnetization to overcome the energy barrier that is separating the two stable magnetization states. This leads to an effective reduction of the coercive field at finite temperatures, which depends on the time scale of the experiment. The detailed knowledge of coercivity as a function of measurement time $t$ and temperature is important for extracting the relevant magnetic properties (such as the zero temperature coercivity, i.e. the anisotropy $H_K$, thermal stability ratio $\beta = KV/(k_B T)$) out of experimental data. Sharrock \cite{shar} gave an analytical expression for a pulsed field experiment 
\beq H_C(t) = H_K\left[1 - \sqrt{\frac{1}{\beta}\ln(f_0 t)}\right] \label{sharrock}\eeq 
and others (\cite{chant}, \cite{el-h}, \cite{pr}, \cite{fv}) have also succeeded in deriving expressions for experiments in which an external field is swept at a constant rate. However, to our best knowledge, there is so far no explicit analytical derivation of the probability density function (PDF) describing the switching field distribution (SFD) arising from the presence of thermal fluctuations exclusively. Knowing the SFD and its relevant parameters is of great importance for the optimization of recording media as well as sensing technologies for magnetic fields as described in \cite{breth}.\\
According to the N\'eel-Brown model \cite{neel}, \cite{brown} the magnetization's spacial orientation fluctuates due to random magnetic fields present at finite temperature. This results in thermal instability and forces the magnetization to switch between its two stable orientations separated by an energy barrier $\Delta E$ at a rate
\beq f = f_0 \exp\left( - \frac{\Delta E}{k_B T}\right) \label{anl}\eeq

Eq. \ref{anl} is known as the Arrhenius-N\'eel law. The preexponential factor $f_0$ is called the attempt frequency at which the magnetization tries to switch orientation and is usually treated as a constant. However, as already shown by Brown \cite{brown}, even for the very simple case of a single domain particle with a field $H$ applied parallel to its easy axis, the attempt frequency (in Hz) takes the form
\beq f_0 = \frac{\alpha \gamma}{1+\alpha^2}\sqrt{\frac{H_K^3 J_S V}{2 \pi k_B T}}\left(1-\frac{H}{H_K}\right) \left(1-\frac{H^2}{H_K^2}\right)\label{att_freq}\eeq 

where $\alpha$ is the damping parameter from the Landau-Lifshitz-Gilbert equation, $\gamma = \gamma_e/\mu_0 = 2.21\cdot10^5$~m/(A~s) is the electron gyromagnetic ratio divided by the permittivity, $H_K$ in amperes per meter is the anisotropy field, $J_S$ in Tesla is the saturation magnetization, $k_B = 1.38\cdot10^{-23}$~J/K is Boltzmann's constant and $T$ in Kelvin denotes the temperature. \\
In the first part, following the work of Kurkij\"arvi \cite{kurki} we will derive the switching field distribution from a master equation and the Arrhenius-N\'eel Law. In the second part, we will compare the analytical result to the output of a Monte-Carlo simulation written in MATLAB. The simulation allows us to introduce a function, which computes the attempt frequency according to eq.~\ref{att_freq} in every time step. This way it is possible to compare the data from the Monte-Carlo simulation to a Langevin dynamics simulation based on the Landau-Lifshitz-Gilbert equation. \\
Furthermore, we will discuss the rate dependence of the coercivity and its standard deviation and will compare the results to the models described in references \cite{chant} - \cite{fv}. In the conclusion we will address the open questions considering the validity of the N\'eel-Brown model, which are closely connected to the understanding of the transition regime between thermally activated and dynamic switching of a single domain particle.
%Specifically, for a magnetic field sensor as described in ref the knowledge of the exact pdf will be of interest in the context of signal processing for noise reduction. The sensor uses a magnetic tunnel junction as the sensing element, which is periodically switched along its easy axis from its high to low resistance state. A low magnetic field applied parallel to the easy axis shifts the switching field of the free layer and is sensed similarly as in a fluxgate magnetometer (ref) in the second harmonic of the alternating resistance signal. Because the field that has to be measured is determined via the rate-dependent switching field of the free layer detailed knowledge of the sfd will be crucial for the understanding of the detection limit. The details of this specific problem will be subject to further work. In this paper we will discuss our results for a parameterset typical for perpendicular magnetic recording media. \\

\section{Analytical model}
The energy landscape for the magnetization of a single domain particle with an external magnetic field $H$ applied parallel to its easy axis has two stable minima separated by a barrier \cite{sw}
\beq \Delta E = \frac{KV}{k_B T}\left(1 - \frac{H}{H_K}\right)^2 \label{deltaE} \eeq
 where $K$ is the anisotropy constant and $V$ the volume. There exist similar expressions for more complex reversal paths of the magnetization, which essentially differ in the exponent. The exact shape of the energy barrier can be calculated using for example the nugded elastic band method \cite{suess}. \\
 When the field $H(t)$ is ramped up, the time dependent probability $P_{\mrm{not}}$ that the particle has not switched until a certain moment $t_0$, is described by a master equation:
\beq \frac{dP_{\mrm{not}}}{dt} = -f P_{\mrm{not}} \label{master}\eeq
where f is the transition rate from the Arrhenius-N\'eel Law given in eq. \ref{anl}. It has to be pointed out here that the Arrhenius-N\'eel Law only applies in the limit of $\Delta E > k_BT$ as originally stated by Kramers \cite{kramers} within transition state theory.

% $f_0$ is the attempt frequency in Hz at which the particle's magnetization tries to overcome the barrier $\Delta E$ and is a function of the magnetic material parameters as well as the external magnetic field. According to Brown (ref) $f_0$ can be calculated by 
%\beq f_0 = \frac{\alpha \gamma}{1+\alpha^2}\sqrt{\frac{H_K^3 J_S V}{2 \pi k_B T}}\left(1-\frac{H}{H_K}\right) \left(1-\frac{H^2}{H_K^2}\right)\label{att_freq}\eeq 
%for an axially symmetric particle with the field applied along the symmetry axis. However, in the following we will regard $f_0$ as constant with respect to the external field, as this is also the case for the other models of time dependent coercivity  (refs) considered here. 
From eq.~\ref{master} we get the following expression for $P_{\mrm{not}}$
\beq \ln P_{\mrm{not}} = - \int_{-\infty}^{t_0} f_0 \exp\left(- \frac{\Delta E}{k_B T}\right) dt \label{lnP}\eeq
as also derived in \cite{fv}. To solve the integral on the right hand side of eq.~\ref{lnP} we use the substitution
\[ u = \sqrt{\beta}\left(1-\frac{H(t)}{H_K}\right) \qquad \beta = \frac{K V}{k_B T}\]
and 
\[H(t) = R t\]
where $R$ is the field rate for a linearly swept field in T/s. 
The result of eq.~\ref{lnP} is then
\beq - \int_{u_0}^{\infty} \frac{f_0 H_K}{\sqrt{\beta} R} \exp(- u^2) du = - \frac{f_0 H_K}{2 R}\sqrt{\frac{\pi}{\beta}}\left[1-\mrm{erf}(u_0)\right]\eeq
and the probability of switching is given by
\beq P(u_0) = \exp\left\{- \frac{f_0 H_K}{2 R}\sqrt{\frac{\pi}{\beta}}\left[1-\mrm{erf}(u_0)\right]\right\} \quad \mrm{with} \quad u_0 = \sqrt{\beta}(1- H_C/H_K)^2 \label{P}\eeq
$P(u_0(H_C))$ is the cumulative distribution function (CDF) which describes the likelihood for the particle to have switched at a field $H$. The switching field distribution (SFD) is then given by the probability density function (PDF) which is the derivative of eq.~\ref{P}
\beq \frac{dP}{dH} = \frac{f_0}{R} \exp\left\{- \frac{f_0 H_K}{2R}\sqrt{\frac{\pi}{\beta}}\left[1-\mrm{erf}(u_0)\right]\right\}\cdot \exp(-u_0^2)  \label{sfd}\eeq

\section{Results}
\subsection{Switching field distribution}
\begin{figure}[h]
\subfigure[Simulation data and analytical result for the switching field distribution, $f_0 = const.$]
{
\resizebox{0.8\textwidth}{!}{\includegraphics{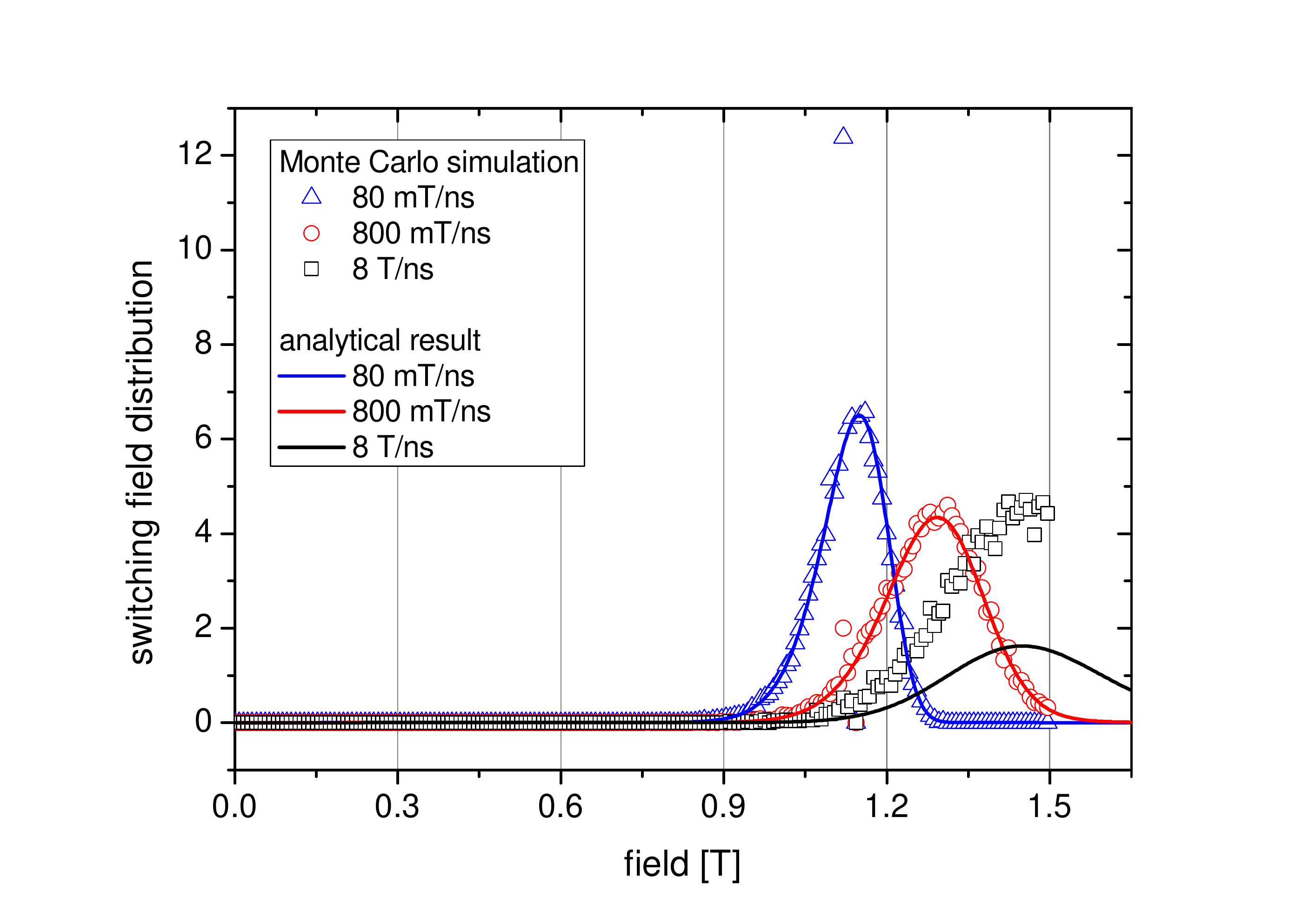}}
}
\label{sfdmc}
\subfigure[Monte-Carlo simulation with $f_0$ computed according to eq.~\ref{att_freq} in every field step and results from a Langevin dynamics simulation ($\alpha=$ 0.02). The solid lines are fits using eq.~\ref{sfd} with $f_0$ as the fit parameter.]
{
\resizebox{0.8\textwidth}{!}{\includegraphics{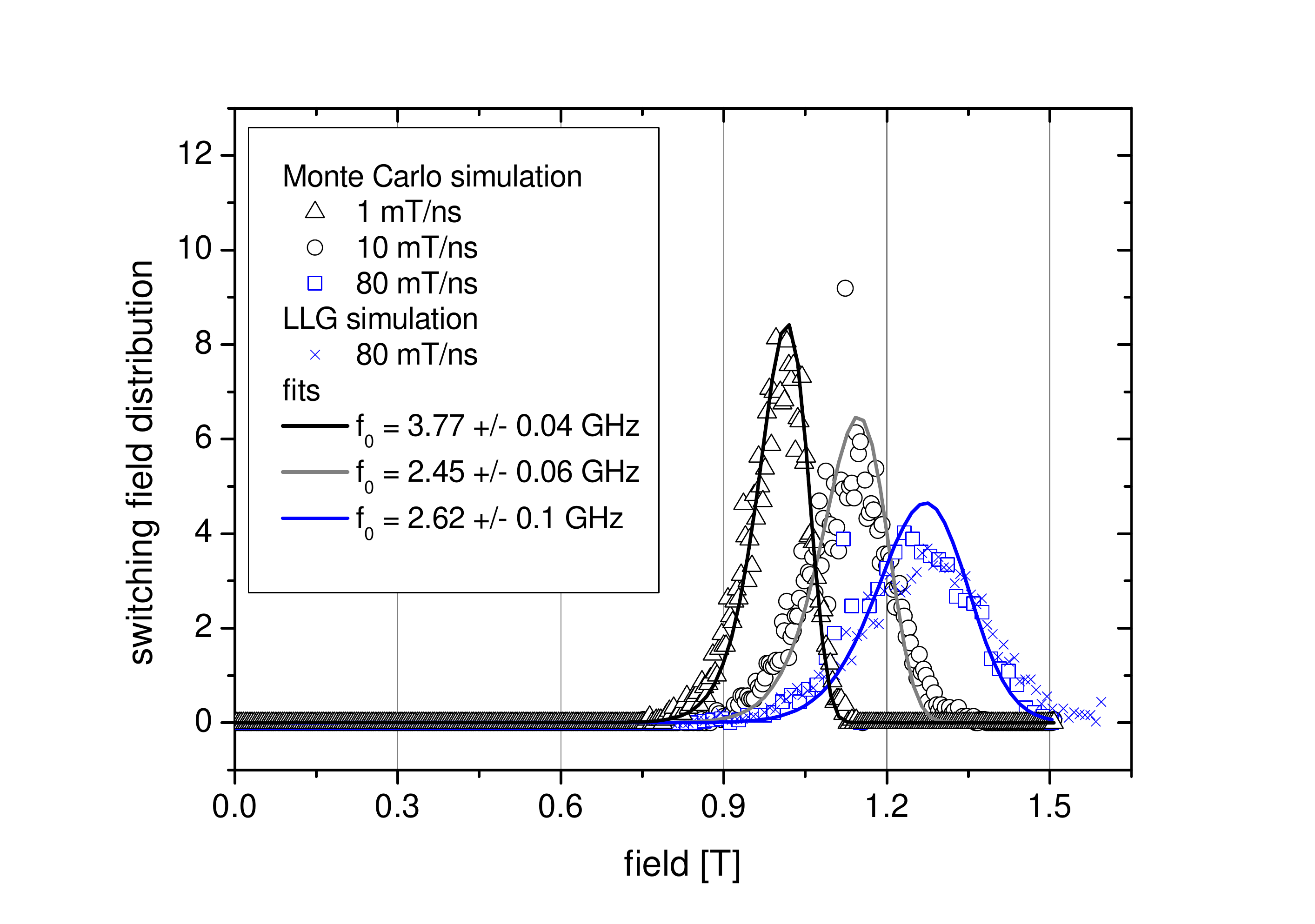}}
}
\label{sfdf0}
\caption{Switching field distributions at different field rates for $K = 0.3$~MJ/$\mrm{m}^3$, $V = (8.8\mrm{nm})^3$, $J_S=$~0.5~T, $\mu_0 H_K =$~1.508~T}
\end{figure}

In fig. \ref{sfdmc} the analytical SFD from eq.~\ref{sfd} is plotted together with the results of a Monte-Carlo simulation written in MATLAB. We assume a single macrospin switching between two magnetization states (up=~+1, down =~-1) in an external field $H(t)$ swept at a constant rate. When the external field is varied as a function of time the switching probability is calculated by \cite{suess}
\beq P = \exp\left\{- \Delta t f_0 \exp\left[-\beta \left(1- \frac{H(t_i)}{H_K} \right)^2\right]\right\} \eeq
where $\Delta t$ denotes the time step of the simulation and $H(t_i)$ is the value of the swept field at a certain time $t_i$. Following the Metropolis algorithm, the values of $P$ are then compared to a random number $x \in [0;1]$. The value of $H$ gets accepted as a switching field $H_C$ as soon as $x > P$. At a value $H(t) \geq H_K$ the magnetization is automatically switched. We performed 100 sweep cycles and then computed the histogram of the obtained values for the switching field. The material parameters applied in the simulation are typical for modern perpendicular magnetic recording materials. $f_0$ was calculated using eq.~\ref{att_freq} for zero field.\\

We can see, that the simulation data agree very well with the analytical result from eq.~\ref{sfd} for the lower field rates. At high rates, however, a large discrepancy occurs between the analytical result and the simulation. If we compute the integral over the analytical SFD for 8~T/ns we get a value $< 1$, which contradicts the general property of a PDF that
\[ \int_{-\infty}^{\infty} \frac{dP}{dH} dH = 1 \]
This implies that the CDF defined by eq.~\ref{P} must only give values between 0 at $-\infty$ and 1 at $+\infty$. For  $\mrm{erf}(+\infty) = 1$ we get
\[\lim_{u_0 \to +\infty} P(u_0) = \exp\left[-\frac{f_0 H_K}{2 R}\sqrt{\frac{\pi}{\beta}}(1-1)\right] = 1 \]
but for $\mrm{erf}(-\infty) = -1$ the normalization condition
\[\lim_{u_0 \to -\infty} P(u_0) = \exp\left[-\frac{f_0 H_K}{R}\sqrt{\frac{\pi}{\beta}}\right] = 0\]
 is only fulfilled if 
\beq R << \sqrt{\frac{\pi}{\beta}} f_0 H_K \label{ratelim} \eeq
For the parameter system discussed here the right hand side of eq.~\ref{ratelim} gives a value of 7.2~T/ns, which explains why eq.~\ref{sfd} fails as a model for the simulation data at a rate of 8~T/ns as shown in fig.~\ref{sfdmc}. \\
In fig. \ref{sfdf0} we present results from a micromagnetic simulation solving the Landau-Lifshitz-Gilbert-equation with an additional random thermal field \cite{langevin} together with a modified Monte Carlo simulation, where the attempt frequency $f_0$ is now calculated in every field step according to eq.~\ref{att_freq}. To keep computation times resonable 80~mT/ns was the lowest rate that could be simulated within the micromagnetic approach. We see good agreement of the two simulation data sets (blue), however, fitting the analytical expression for the SFD from eq.~\ref{sfd} does not work satisfyingly. Fitting the data from the Monte Carlo simulation works better for lower rates, which is indicated by the decreasing standard deviation of the fit parameter $f_{0,\mrm{eff}}$, which we call the effective attempt frequency. Compared to the model in fig.~\ref{sfdmc} where $f_0$ was set constant to $f_0(H=0 \mrm{T}) = 21.1$~GHz we see a shift of the mean switching field to higher values, which is related to the lower effective attempt frequency  we get from the fit for which $f_0(H = H_C) < f_{0,\mrm{eff}} < f_0(H = 0)$ applies.

\subsection{Rate dependence of the coercivity}
The coercivity of a magnetic recording medium is of special interest regarding the write process of a magnetic bit.  Coercivity measurement methods usually employ very low field rates (e.g. VSM: $<$1~T/s) in contrast to the high field rates of up to 1~T/ns used in modern hard disk drives for the write process. Various models have been proposed to extrapolate between the laboratory time scale and the actual time scale used in magnetic recording. An overview of the expressions derived in references \cite{chant} - \cite{fv} is given in table \ref{coercmodels}.

\begin{table}
\begin{ruledtabular}
\begin{tabular}{ll}

Chantrell \textit{et al.} \cite{chant} & $R(H_C) = H_K f_0 \exp[-\beta(1 - H_C/H_K)^2]/[2 \beta(1-H_C/H_K)]$	\\
Feng and Visscher \cite{fv} & $R(H_C) = (2 \ln2 \beta)^{-1} \pi f_0 H_K \left\{1 - \mrm{erf}[\sqrt{\beta}(1-H_C/H_K)]\right\}$\\
El Hilo \textit{et al.} \cite{el-h} & $H_C(R) =  H_K\{1 - \sqrt{\beta^{-1}\ln[f_0 H_K/(2 R \beta)]}\}$\\
Peng and Richter \cite{pr} & $H_C(t_{\mrm{eff}}) =  H_K\{1 - \sqrt{\beta^{-1}\ln[f_0 t_{\mrm{eff}}/(2 \ln2)]}\} \quad t_{\mrm{eff}} = t_{\mrm{eff}}(R, \beta)$\\

\end{tabular}
\end{ruledtabular}
\caption{Analytical expressions for rate-dependant coercivity}\label{coercmodels}
\end{table}

\begin{figure}[h]
\resizebox{0.8\textwidth}{!}{\includegraphics{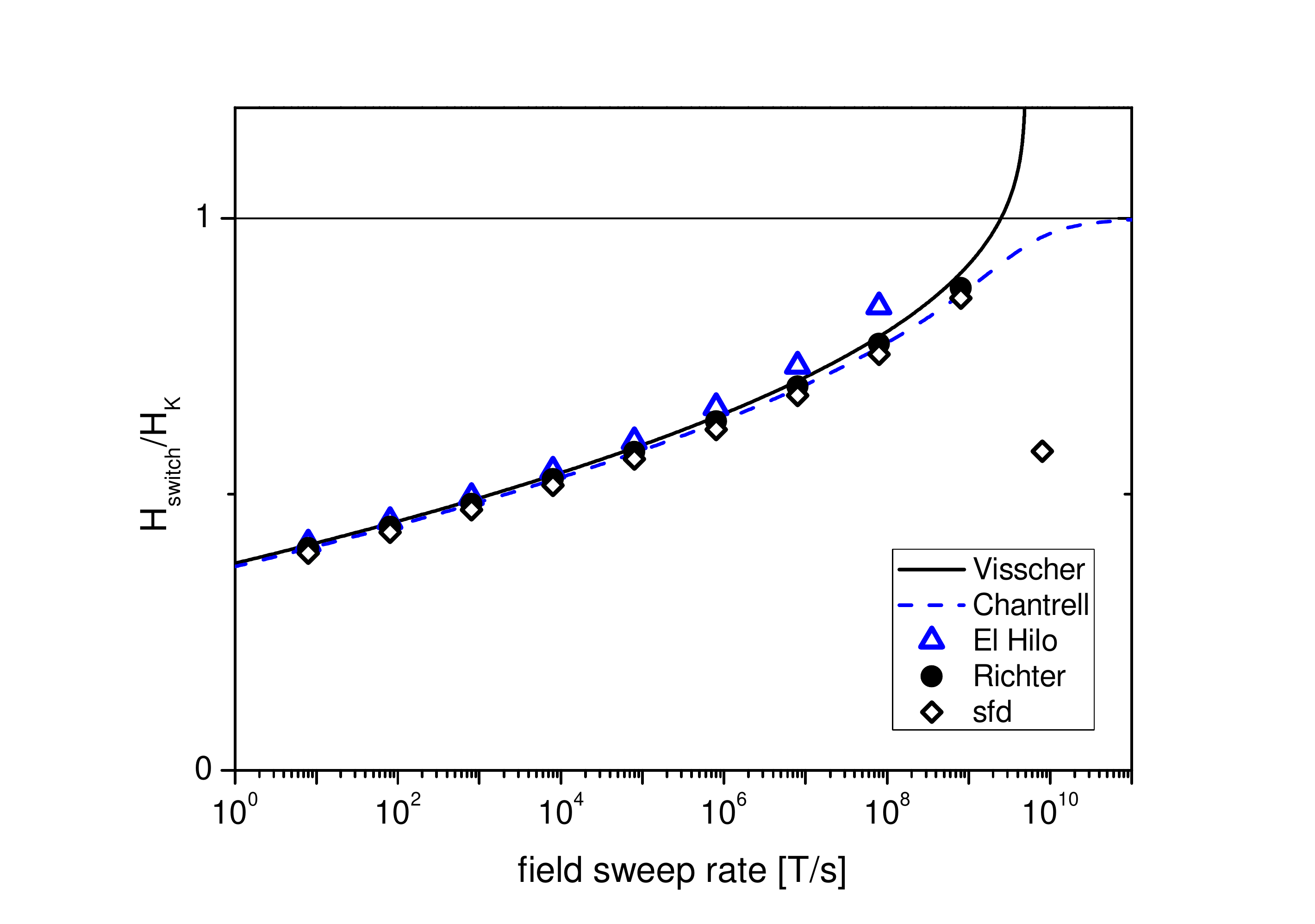}}
\caption{Comparison of the models from table~\ref{coercmodels} and eq.~\ref{mean} for the rate-dependent coercivity}\label{modelsgraph}
\end{figure}

We calculate the switching field directly from the SFD. The mean value $\overline{H}$ of the variable H with the PDF $dP/dH$ is given by
\beq \overline{H} = H_C = \int_{-\infty}^{\infty} H \frac{dP}{dH} dH \label{mean}\eeq
Because $dP/dH$ is not an analytical function we have to use numerical integration to calculate $\overline{H}$. As it is shown in fig. \ref{modelsgraph} the switching fields derived this way are in good agreement with the other models, which have proven to adequately describe realistic recording media up to a certain field rate. We observe, however, that the models strongly deviate from each other at approx. 1~T/ns, which is a typical rate used in magnetic recording. As already stated above, the limit for failure of our model is determined by the loss of normalization of the PDF (see eq.~\ref{ratelim}). Peng and Richter \cite{pr} give $R << f_0 H_K (2\ln2 \sqrt{\beta})^{-1} = $ 2.9~T/ns as a condition for the validity of their model, which is a similar expression to the limit we derived in eq.~\ref{ratelim}. The expressions by El Hilo \textit{et al.} \cite{el-h} as well as Peng and Richter \cite{pr} give non-real values as soon as the rates are out of the range of validity for the assumptions made to derive their expressions. \\

\subsection{Rate dependence of the standard deviation}
The standard deviation of the switching field can be calculated from the second moment of the PDF, i.e. the variance:
\beq \sqrt{\mrm{var}(H)} = \sigma =\sqrt{\int_{-\infty}^{\infty} (H - \overline{H})^2 \frac{dP}{dH} dH }\label{stddev}\eeq
\begin{figure}[h]
\resizebox{0.8\textwidth}{!}{\includegraphics{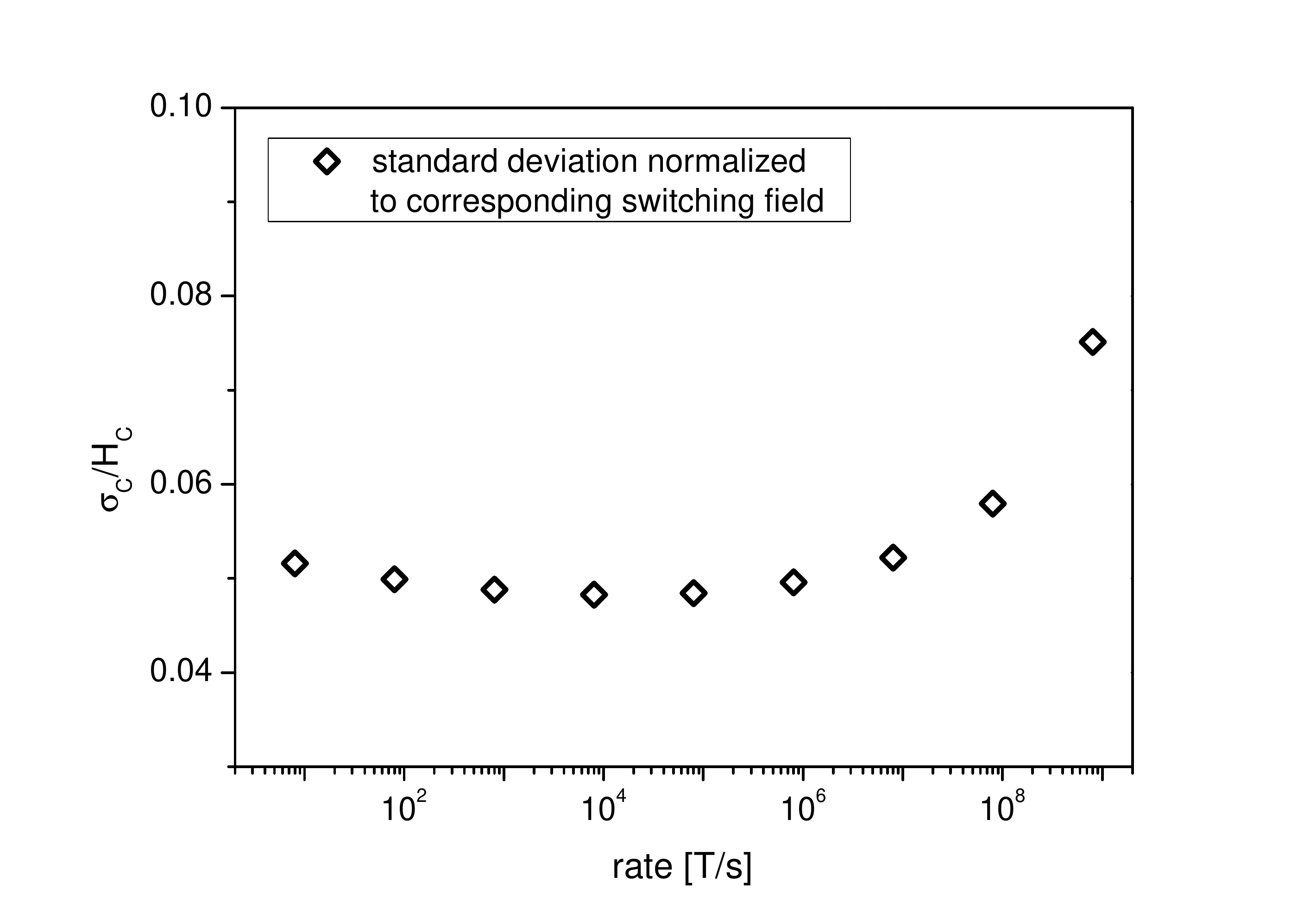}}
\caption{Rate dependence of the standard deviation of the switching field} \label{stddevgraph}
\end{figure}
Fig. \ref{stddevgraph} shows the standard deviations $\sigma_C$ normalized to the switching fields $H_C$ plotted with respect to the field rate. It is interesting to see that $\sigma_C/H_C$ first decreases slightly, but then increases at higher field rates. Most importantly, $\sigma_C/H_C$ has a minimum finite value at an intermediate field rate.  
We like to emphasize that the standard deviation in eq.~\ref{stddev} is derived for thermal fluctuations acting on a single domain particle, which would correspond for example to a single grain in a perpendicular recording medium. This is an entirely different effect as the standard deviation described by Hovorka \textit{et al.} \cite{hovo}, which arises from anisotropy and volume distributions of the magnetic medium. Hence, a direct comparison of the values for $\sigma_C/H_C$ to values from the relevant literature is not reasonable. We are convinced, however, that the standard deviation derived here represents a fundamental limit regarding the accurate switching of nanoscale magnetic particles.

%We see that the standard deviation for the highest field rate strongly deviates from the other values. We can understand this increase when looking at the sfd in fig. \ref{sfdgraph}. However, we have argued before, that values calculated for rates larger than 7.2~T/ns (as given by eq.~\ref{ratelim}) can not be accurately described anymore by our model. To get a better picture of the trend of the standard deviation with respect to the corresponding mean switching field at lower field rates we have plotted the values $\sigma / H_{\mrm{switch}}$ in the inset of fig. \ref{stddevgraph}. The values of $\sigma / H_C$ differ from the values calculated for example by Hovorka et al. (ref) by the use of the $\Delta H(M, \Delta M)$ method in their magnitude. The most striking difference is that for our model $\sigma / H_{\mrm{switch}}$ first decreases slightly and then increases at higher field rates. In the $\Delta H(M, \Delta M)$ method a lognormal distribution is used to model the sfd and then the standard deviation is derived as a fitting parameter. 

\section{Conclusion}

We described an entirely analytical approach to derive the SFD of a single-domain particle caused by thermal fluctuations. Unlike the SFDs derived from an underlying distribution of grain sizes and switching volumes, the distribution presented here originates from the thermal activation of the particle's magnetization for overcoming an energy barrier at a rate described by the Arrhenius-N\'eel Law. The effect exists independently of any other contribution to the SFD and therefore represents a fundamental limit to the accuracy of magnetization reversal at finite temperatures. An experimental evidence of the validity of the N\'eel-Brown model at low field rates was given by Wernsdorfer \textit{et al.} \cite{werns}. For typical field rates of $> 0.1$~T/ns used in magnetic recording micromagnetic simulation tools based on solving the Langevin equation describing dynamic switching under the influence of thermal fluctuations are used to compute the switching fields. However, computation time increases dramatically towards lower field rates, where thermal switching is dominant. It will be subject to future work to investigate this transition regime and understand the underlying physical processes. \\
Furthermore, we are able to give an upper limit for the field rates where the assumptions of the N\'eel-Brown model are applicable. As with increasing field rates the switching fields approach the value of zero temperature coercivity, we show failure of the model due to the loss of normalization of the PDF describing the SFD. Using a Monte Carlo simulation for a single macrospin we reproduced the result of the analytical model within its range of validity and we were also able to include the field-dependence of the attempt frequency for the transition rate. By doing so, we observe significant changes to the switching field. We conclude, that the usual approach of taking the attempt frequency as a constant does not adequately describe rate-dependent coercivity, which is additionally supported by Langevin dynamics simulation data. 

%We described an entirely analytical approach to derive the sfd of single-domain particle caused by thermal fluctuations. Instead of simulating an ensemble of non-interacting particles with a distribution of volumes and anisotropies we switch one particle several times using a Metropolis algorithm and plot the histogram of the switching fields. We see very good agreement with the analytical expression (eq.~\ref{sfd}) and other models for rate-dependent coercivity up to a certain field rate at which the integral over the sfd becomes unnormalized. Based on this, we consider the assumptions made for the validity of the Arrhenius-N\'eel model to be put under question, as at high field rates as $H_C \to H_K$ the energy barrier (eq.~\ref{deltaE}) cannot be regarded as high with respect to the thermal energy $k_B T$. Furthermore, the influence of the attempt frequency needs to be investigated, as it is strongly dependent on the external field. On the nanosecond timescale, which is the limit defined in eq.~\ref{ratelim}, the regime of dynamic magnetization switching is entered, which can be treated in the frame of micromagnetism by solving the Langevin-equation (refs). Investigating this transition regime will be subject to future work.

\end{document}